\documentclass[aps,prd,reprint,preprintnumbers,nofootinbib]{revtex4-1}

\usepackage{amsmath}
\usepackage{amsfonts} 
\usepackage{amssymb}
\usepackage{graphicx}
\usepackage{hyperref}
\usepackage{tensor}
\usepackage{siunitx}
\usepackage{float}
\usepackage{bm}
\usepackage{mathtools}

\newcommand{\be}{\begin{equation}}
\newcommand{\ee}{\end{equation}}

\begin{document}

\title{\Large Emergent potentials and non-perturbative open topological strings}

\author{\large Jarod Hattab}
\author{\large Eran Palti}

\affiliation{\vspace{0.4cm}
Department of Physics, Ben-Gurion University of the Negev, Beer-Sheva 84105, Israel}

\begin{abstract}
We show that integrating out M2 branes ending on M5 branes inside Calabi-Yau manifolds captures non-perturbative open topological string physics. The integrating out is performed using a contour integral in complexified Schwinger proper time. For the resolved conifold, this contour can be extended to include the zero pole, which we argue captures the ultraviolet completion of the integrating out and yields the tree-level polynomial terms in the free energy. This is a manifestation of the Emergence Proposal, and provides further evidence for it. Unlike the case of closed strings, where the emergent terms are kinetic terms in the action, for these open strings it is tree-level potential terms which are emergent. This provides a first quantitative example of the proposal that classical tree-level potentials in string theory emerge from integrating out co-dimension one states.
\vspace{1cm}
\end{abstract}

\maketitle

\section{Introduction}
\label{sec:int}

Recently, in \cite{Hattab:2024ewk}, it was shown that integrating out M2 branes on Calabi-Yau manifolds yields the non-perturbative free energy for closed topological strings. The integrating out is similar to the seminal calculation of Gopakumar and Vafa \cite{Gopakumar:1998ii,Gopakumar:1998jq}, but differs quantitatively. Specifically, it must be performed as a contour integral in complex Schwinger proper time. The Ooguri-Vafa formula \cite{Ooguri_2000} is the open-string extension of the Gopakumar-Vafa formula, integrating out M2 branes ending on M5 branes. In this paper, we show that it similarly should be evaluated using a contour integral, and again this captures non-perturbative open topological string physics. 

The contour integral formulation of the integrating out procedure also allows for an evaluation of the ultraviolet modes contribution of the integrated out states, so the removal of the ultraviolet cutoff in the Schwinger integral. In some special settings, specifically for our case the resolved conifold Calabi-Yau, the ultraviolet completion is captured simply by extending the integration contour to include the pole at the origin \cite{Hattab:2023moj}. Including the ultraviolet contribution this way yields additional terms in the free energy. In this paper, we show that these are precisely the tree-level classical terms of the effective action for open strings. In analogy with a similar result for closed strings \cite{Hattab:2023moj}. This is a manifestation of the Emergence Proposal \cite{Palti:2019pca}, which was based on the ideas in \cite{Harlow:2015lma,Heidenreich:2017sim,Grimm:2018ohb,Heidenreich:2018kpg,Palti:2019pca}. It states that the dynamics of all fields arises due to integrating out towers of states. The proposal has been studied in some detail within string theory, see \cite{Grimm:2018cpv,Corvilain:2018lgw,Lee:2019xtm,Lee:2019wij,Palti:2020tsy} for a sample of some early work, and  \cite{Marchesano:2022avb,Castellano:2022bvr,vandeHeisteeg:2022btw,Cota:2022maf,Cribiori:2022nke,Marchesano:2022axe,Blumenhagen:2023tev,Castellano:2023qhp,Burgess:2023pnk,Baume:2023msm,Cribiori:2023ffn,Blumenhagen:2023yws,Blumenhagen:2023xmk,Seo:2023xsb,Calderon-Infante:2023ler,DeBiasio:2023hzo,Calderon-Infante:2023uhz, Castellano:2023aum,Castellano:2023jjt,Castellano:2023stg,Marchesano:2023thx,Basile:2023blg,Cota:2023uir,Hattab:2023moj,Casas:2024ttx,Blumenhagen:2024ydy,Hattab:2024thi,Blumenhagen:2024lmo,Hattab:2024ewk}, for more recent investigations.

Interestingly, for the resolved conifold, including the zero pole for the closed string sector induces tree-level kinetic terms. For the open-string sector, it induces tree-level potential terms. One way to understand this is to think about the M5 brane world-volume which spans only two non-compact dimensions. In this sense, it is a two-dimensional theory, and integrating out particles (with a one-dimensional worldline) then amounts to integrating out co-dimension one objects. In terms of emergence, it was proposed already in \cite{Palti:2019pca} that tree-level potentials in string theory should be emergent from integrating out co-dimension one objects. The results of this work are therefore evidence for this idea.

While our quantitative results here only apply to the resolved conifold, we also comment in section \ref{sec:dualemer} about how to possibly generalize them to other Calabi-Yau manifolds. We frame this discussion in the context of a specific picture of emergence that we term Dual Emergence.

\section{Open topological strings}
\label{sec:ops}

In \cite{Hattab:2024ewk} a formula for integrating out wrapped M2 branes was presented which was closely tied to the Gopakumar-Vafa (GV) formula \cite{Gopakumar:1998ii,Gopakumar:1998jq}. The Ooguri-Vafa (OV) formula is a similar formula to GV, but which accounts for an open string sector \cite{Ooguri_2000}. In this section we propose the analogous modification of it, which we claim captures non-perturbative open-string physics (though not necessarily all such physics). 

In \cite{Ooguri_2000} Ooguri and Vafa considered a stack of D4 branes with support on $\mathbb{R}^2\subset \mathbb{R}^4$ and wrapping a Lagrangian submanifold $L\subset Y$, in a Calabi-Yau manifold $Y$. For simplicity, henceforth we consider only a single such D4 brane.\footnote{This can easily be extended to $N$ coincident D4-branes and multiple Lagrangian submanifolds $L_i$ as explained in \cite{Ooguri_2000,Dedushenko:2014nya}.} The worldvolume action of the D4 contains certain F-terms, which can be written as
\begin{equation}
\label{target}
    I_n = \int \text{d}^2x\;\text{d}^2\theta \mathcal\,{R}_n(\mathcal{X}^{I},\mathcal{U}^{\rho})\left(\mathcal{W}_{||}\right)^n \;.
\end{equation}
Here $\mathcal{X}^{I}$ and $\mathcal{U}^{\rho}$ are chiral superfields in the theory with $(2, 2)$ supersymmetry on $\mathbb{R}^2$ whose bottom components, $t^i$ and $u^\rho$, are complex scalars. The $t^i$ are the usual closed-string sector fields, composed of the Kahler moduli and the $B$-field. The $u^{\rho}$ are counted by $b^1(L)$, and are comprised of the geometric moduli of $L$ as well as the Wilson lines from the brane gauge field. Note that the splitting between the $t^i$ and $u^{\rho}$ is somewhat arbitrary, and different components can be moved between them. Similarly, $\mathcal{W}_{||}$ is a chiral superfield whose bottom component is the anti-self-dual part of the graviphoton field strength $W_{||}^-$ along the $\mathbb{R}^2$. See \cite{Ooguri_2000,Marino:2005sj,Dedushenko:2014nya} for more details.

In terms of topological open strings, the $R_n$ are associated to contributions to the topological string free energy $R$, coming from maps with with genus $g$ and $h$ boundaries on $L$, such that $n = 2g+h-1$ \cite{Ooguri_2000,Marino:2005sj}. 

Lifting the setting to M-theory, the sum $\sum_{n = 0}^{+\infty}I_n$ was reinterpret in \cite{Ooguri_2000} as the asymptotic expansion of the superspace effective action obtained from integrating out, in Schwinger's formalism, specific BPS particle excitations of wrapped M2 branes ending on the M5 brane wrapping the Lagrangian submanifold $L$ and the M-theory circle in a constant anti-selfdual graviphoton background.

The integrating out calculation was argued to yield an expression for the exponential parts of the perturbative (in $\lambda$) open topological string free energy $R^p$, or equivalently the F-terms on the support of the D4 brane, if we identify $\lambda = g_sW^{-}_{||}$ with $g_s$ the type IIA string coupling \cite{Ooguri_2000,Dedushenko:2014nya,Marino:2005sj}
\be 
\label{Rp}
R^{p} = \sum_{q, n} \int_{\epsilon}^{\infty}\frac{\text{d}\tau}{\tau}\frac{\mathrm{Tr}_{{\cal H}_{q}}\left(-1\right)^{F}e^{is\lambda \tau}e^{-z_{q,n}\tau}}{2\sin(\tau\lambda/2)}\;.
\ee
Here $\mathrm{Tr}_{{\cal H}_q}\left(-1\right)^{F}$ is the sum over the BPS state Hilbert space, weighted by $F=S_L + S_R$ where $S_{L,R}=S\pm R$ with $S$ being the two-dimensional spin and $R$ the R-symmetry $U(1)$ charge. This includes the degeneracy factors, the analogs to the Gopakumar-Vafa invariants. We also have $s=2S_L$. As usual, $\epsilon$ is an ultraviolet cutoff.

The BPS states are counted geometrically by the relative homology $H_2(Y,L;\mathbb{Z})$ which is parameterised by charges $q = (\beta_i,r_{\rho})$. The central charge is then
\be 
z_{q,n} = \beta\cdot t+r\cdot u-2\pi in \;,
\ee 
with KK number $n$ along the M-theory circle.

Following the steps of \cite{Hattab:2024ewk}, we propose that instead of (\ref{Rp}) we should consider 
\be
\label{R}
R = \sum_{q, n}\int_{0^+}^{\infty}\frac{\text{d}\tau}{\tau}\frac{\mathrm{Tr}_{{\cal H}_{q}}\left(-1\right)^{F}e^{is\lambda \bar{z}_{q,n}\tau}e^{-|z_{q,n}|^2\tau}}{2\sin(\bar{z}_{q,n}\tau\lambda/2)}\;.
\ee
And that this gives not only perturbative contributions, but also non-perturbative ones. 

Following the same steps as \cite{Hattab:2024ewk} this can be evaluated for $\lambda\in \mathbb{R}^+$ to give
\be
\label{full}
R = \sum_{q}\oint_{C}\frac{\text{d}w}{w}\frac{\mathrm{Tr}_{{\cal H}_{q}}\left(-1\right)^{F}e^{is\lambda w}}{1-e^{-2\pi i w}}\frac{e^{-(\beta\cdot t+r\cdot u-2\pi i n_q)w}}{2\sin(w\lambda/2)}\;,
\ee
where $C$ is the anti-clockwise closed contour in the complex $w$-plane, that starts at $0^+$ encircling all the poles on the real axis and 
\be
n_q = \Bigl\lfloor\beta\cdot\text{Im}\left(\frac{t}{2\pi}\right)+r\cdot\text{Im}\left(\frac{u}{2\pi}\right)\Bigr\rfloor\in\mathbb{Z}\;.
\label{nqdef}
\ee
The claim is that (\ref{full}) captures topological open string non-perturbative effects, associated to the poles at
\be 
w \in \left(\frac{2\pi}{\lambda}\right) \mathbb{N}^* \;.
\ee
It is not clear to us if this captures all the non-perturbative physics associated also to topological open strings, but it certainly captures a certain class of such non-perturbative effects. 

\section{The resolved conifold and the ultraviolet completion}

The integrating out expression (\ref{full}) has an ultraviolet cutoff at $0^+$, so arbitrarily close to the zero pole, but not including it. It was argued in \cite{Hattab:2023moj} that, for closed strings, in the special case of the Calabi-Yau being the non-compact resolved conifold, one can complete the integrating out in the ultraviolet by simply including the zero pole in the contour. More generally, so for other Calabi-Yau manifolds, we do not expect that this is a correct ultraviolet completion, and we discuss this issue in section \ref{sec:dualemer}.

We henceforth restrict to the case of the resolved conifold, and to the proposed ultraviolet completion. We first discuss the closed string case in section \ref{sec:resconclo}, and then the open string case in section \ref{sec:resconop}. The ultraviolet completion of the integrating out calculation is particularly interesting in the context of the emergence proposal \cite{Palti:2019pca}. Specifically, as discussed in section \ref{sec:dualemer}, it is conjectured that the ultraviolet contribution to the integrating out yields the classical tree-level terms in the effective action \cite{Hattab:2023moj}. In this section we show that this is the case for both the closed string and open string cases. 

\subsection{Kinetic terms for closed strings}
\label{sec:resconclo}

In the pure closed-string case, the resolved conifold has a spectrum of states consisting of a single wrapped D2 brane and all its D0 bound states. The resulting integrating out calculation gives \cite{Hattab:2024ewk}
\be
\label{rcclosed}
    F^{rc} = \oint_C\frac{\text{d}w}{w}\frac{1}{1-e^{-2\pi i w}}\frac{e^{-\left(t-2\pi i n_{b} \right)w}}{\big(2\sin(w\lambda/2)\big)^{2}} \;,
\ee
with $n_b = \Bigl\lfloor\text{Im}\left(\frac{t}{2\pi}\right)\Bigr\rfloor\in\mathbb{Z}
\;$.
We are interested in the low-energy two-derivative type IIA effective action, which is given by the genus zero part $F^{rc}_0$ of (\ref{rcclosed}). In general, the genus zero prepotential $F_0$ determines kinetic terms in the effective action through 
\begin{eqnarray}
\nonumber   -i\int\text{d}^4x\;d^{4}\theta\;F_0(\mathcal{X}^I) +c.c \supset -\int d^4x\;g_{ij}\text{d}t^i\wedge \star \text{d}\overline{t}^j\\
   +\int d^4x\;\left[\frac12\text{Im}\,\mathcal{N}_{IJ} F^{I}\wedge \star F^{J}+\frac12\text{Re}\,\mathcal{N}_{IJ} F^{I}\wedge F^{J}\right] \nonumber\;.
\end{eqnarray}
Here the $t^i$ are the scalars in vector multiplets, the $F^I$ are the field strengths of in those multiplets as well as the graviphoton, $g_{ij}$ is the moduli space metric, and ${\cal N}_{IJ}$ is the gauge kinetic matrix.

If we extend the contour $C$ to include the zero pole we get the polynomial contributions, for $n_b=0$, which are the leading contributions to the prepotential $F_0^{rc}$ \cite{Hattab:2023moj,Hattab:2024ewk}
\be
 -\frac{1}{6}t^3 + \frac{\pi i}{2}t^2 + \frac{\pi^2}{3}t \;.
 \label{zeropoleconclo}
\ee
These match the expected results coming from a geometric transition from Chern-Simons theory \cite{Gopakumar:1998ki}, up to a factor of $\frac12$ (the correct answer is $\frac12$ of (\ref{zeropoleconclo})). As mentioned in \cite{Hattab:2023moj,Hattab:2024ewk}, the factor of $\frac12$ arises because for the resolved conifold the zero pole has a degeneracy of two associated to a flop transition. The negative definite poles in the resolved conifold case should be associated to the flopped manifold with $t' = -t >0$, and the zero pole is shared by both sides. Evaluating the negative poles of $F^{rc}_0$ and half of the zero pole yields
 \begin{eqnarray}
     \frac{t'^3}{12}+\frac{\pi i}{4} t'^2-\frac{\pi^2}{6}t'+\sum_{k \geq 1}\frac{e^{-kt'}}{k^3}\;,
 \end{eqnarray}
 which indeed corresponds to the right rule for the transformation of the polynomials term of $F^{rc}_0$ under a flop ($\kappa = 1/2$ and $c_2 = -1$):
 \begin{eqnarray}
     -\frac{\kappa t^3}{6}&\rightarrow& -\frac{(\kappa-1) t'^3}{6} \;,\\
     -\frac{\pi^2}{6}c_2 t &\rightarrow& -\frac{\pi^2}{6}(c_2+2)t' \;.
 \end{eqnarray} 
 We will see that there is a similar degeneracy issue for the open string case at the zero pole.

\subsection{Potentials for open strings}
\label{sec:resconop}

In the open string case, the relevant leading action is controlled by $R_{0}$. This corresponds to the genus $g = 0$ and boundary $h = 1$. The theory is localised on the $\mathbb{R}^2$ subspace of the external space which coincides with the D4 worldvolume. In this theory, half the supersymmetries are broken and the F-terms associated to $R_{0}$ are part of the superpotential. So, while the relevant prepotential $F_0$ in the closed-string case controlled the kinetic terms for the closed-string fields, here it controls open-string superpotential terms.

In general, the states to be integrated out are D2 branes wrapping two-chains which have a boundary one-cycle on the Lagrangian cycle $L$ wrapped by the D4 brane. For the case of the resolved conifold, there are two types of BPS D2 branes: wrapping the north and south hemispheres of the $S^2$ and intersecting the D4 on the equator (with opposite orientations) \cite{Ooguri_2000}. The Lagrangian cycle $L$ is non-compact, including the equator of the $S^2$ and stretching out over the cone to infinity. 

In terms of the charges introduced in section \ref{sec:ops}, the only two states (with degeneracy one) are of spin-zero $s=0$ and charges $q_+ = (0,1)$ and $q_- = (1,-1)$.
We denote:
\be
    t_+ = u \;\;, \;\;\;\;
    t_- = t-u\;.
\ee
The expression of $R$ for the resolved conifold that we denote $R^{rc}$ is therefore given by
\be
\label{rc}
    R^{rc} = \oint_{C}\frac{\text{d}w}{w}\frac{1}{1-e^{-2\pi i w}}\frac{e^{- (t_+- 2\pi in_+ )w}+e^{- (t_-- 2\pi in_- )w}}{2\sin(w\lambda/2)} \;,
\ee
with $n_{\pm}$ defined as in (\ref{nqdef}).

To calculate the ultraviolet completion of the integrating out we should evaluate the zero pole of (\ref{rc}). This yields (at $n_{\pm}=0$),
\begin{eqnarray}
\label{polypotdeg}
   & & \left[ \frac{1}{\lambda}\left(\frac{t_-^2}{2}-i\pi t_--\frac{\pi^2}{3}\right)+\frac{\lambda}{24} \right] \nonumber \\
     &+& \left[ \frac{1}{\lambda}\left(\frac{t_+^2}{2}-i\pi t_+ -\frac{\pi^2}{3}\right)+\frac{\lambda}{24} \right] \;.
\end{eqnarray}
The two contributions in (\ref{polypotdeg}) are analogous to the factor of two degeneracy in the closed string case, as in (\ref{zeropoleconclo}). We therefore should subtract one of them to account for only one side of the flop. We can then write the final contribution from the ultraviolet as
\be
\label{polypot}
\frac{1}{\lambda}\left(\frac{t_-^2}{2}-i\pi t_--\frac{\pi^2}{3}\right)+\frac{\lambda}{24} \;.
\ee

We would now like to compare (\ref{polypot}) with the tree-level terms in $R^{rc}_0$ and $R^{rc}_1$. Emergence proposes that they should be precisely equal. Because the resolved conifold is non-compact, we cannot determine these by dimensional reduction. Rather, in complete analogy to the closed string case, they can be determined through a geometric transition from a calculation in the dual Chern-Simons theory. This yields \cite{Ooguri_2000}
\be
\label{rcsper}
	R^{rc}_{CS} = \sum_{k\geq 1}\frac{e^{-kt_+}-e^{kt_-}}{2 k\sin(k\lambda/2)} \;.
\ee
The expressions in (\ref{rcsper}) is to be evaluated at $\mathrm{Re\;} t_{\pm} = 0$, which is where the geometric transition takes place. Indeed, it is manifestly not convergent for $\mathrm{Re\;} t_- > 0$. In order to compare it with (\ref{polypot}) we must analytically continue it along $t_{\pm}$. We can do this by first writing it as 
\begin{eqnarray}
\label{rcsper2}
	& &\frac{1}{\lambda} \Big( \mathrm{Li}_2 \left(e^{-t_+}\right) - \mathrm{Li}_2 \left(e^{+t_-}\right)\Big) \nonumber \\ \nonumber
	&+& \frac{\lambda}{24} \Big( \mathrm{Li}_{0} \left(e^{-t_+}\right) - \mathrm{Li}_{0} \left(e^{+t_-}\right)\Big)  + \; {\cal O}\left(\lambda^3\right)\;,
\end{eqnarray}
and then using the natural analytic continuation for poly-logarithms. We then use
\begin{eqnarray}
-\mathrm{Li}_2 \left(e^{+t_-}\right) &=& \mathrm{Li}_2 \left(e^{-t_-}\right) + \frac{(2\pi i)^2}{2}B_2\left(\frac{t_-}{2\pi i}\right) \;, \nonumber \\
-\mathrm{Li}_0 \left(e^{+t_-}\right) &=& \mathrm{Li}_0 \left(e^{-t_-}\right) + 1 \;,
\end{eqnarray}
with $B_2$ the second Bernoulli polynomial. This yields a final expression for the polynomial terms matching precisely (\ref{polypot}). Thus we find the same result as the closed string case, as predicted by emergence.

\section{Integrating out co-dimension one states}
\label{sec:indodim1}

The open string potential arises from integrating out particle states in a two-dimensional theory. The particle worldlines are one-dimensional, and so we are integrating out co-dimension one states to generate a potential. The idea that potentials are emergent from integrating out co-dimension one states was already proposed in \cite{Palti:2019pca}.

One can try to expand the setup so that the potential is a full four-dimensional potential. Indeed, already in \cite{Ooguri_2000} it was proposed how to do this. One replaces the D4 brane with a D6 brane filling all of the non-compact dimensions. The wrapped D2 brane being integrated out is then replaced with a D4 brane, so that it remains co-dimension one. The claim in \cite{Ooguri_2000} is that the resulting exponential terms in the four-dimensional superpotential on the D6 are then just the same as the ones with the D4 calculation, since they are given by the topological string computation which is the same. This then implies the same should hold for the tree-level polynomial terms, associated to the zero pole, studied in this paper.  

If we do not utilize the topological string perspective, then we would need to somehow  integrate out co-dimension one objects directly in the target space. It is not known how to do this in general. However, a nice idea was proposed in \cite{Ooguri_2000} to try and capture such an integrating out. In this section we develop this idea, showing that it also captures the polynomial tree-level pieces.

Let us consider the co-dimension one states as domain walls. We introduce a field $Y_{q,n}$ for each domain wall, such that the vacua on either side of the wall are related by a transformation $Y_{q,n} \rightarrow Y_{q,n} + 2 \pi i$. Then the superpotential is related to the domain wall tension as
\be
R_0(Y_{q,n}-2\pi i )-R_0(Y_{q,n}) = 2\pi i z_{q,n} \;.
\ee
A potential which satisfies this constraint is
\be
R_0 = -z_{q,n}Y_{q,n}-\exp(-Y_{q,n}) \;.
\ee
We can evaluate the minimum of the potential to yield
\be
Y_{q,n} = -\log z_{q,n} \;.
\label{yqnsol}
\ee
Integrating out the domain wall is then captured by integrating out $Y_{q,n}$, which is just replacing it using (\ref{yqnsol}). This yields the potential
\be
R_0 = z_{q,n}\Big(\log z_{q,n}-1\Big) \;.
\ee 
We now need to sum (with degeneracy) over the different domain walls $\sum_{q,n}$. The sum over $n$ is divergent, and must be regularised. There are two ways to do this, the first is by performing the sum directly and then using zeta function regularisation.\footnote{This is exactly the same divergence and regularization as studied in \cite{Blumenhagen:2023tev}.} The second way, physically matching our approach, is to use the integral representation
\be
    z_{q,n}\Big(\log z_{q,n}-1\Big) = \int_{0}^{\infty}\frac{\text{d}\tau}{\tau^2}\left(e^{-z_{q,n}\tau}+z_{q,n}\tau e^{-\tau}-1\right) \;.
    \label{zqnint}
\ee
The regularization is then just the removal of the explicitly divergent last two terms in the integral (\ref{zqnint}). So we set
\be
 z_{q,n}\Big(\log z_{q,n}-1\Big) \rightarrow \int_{0}^{\infty}\frac{\text{d}\tau}{\tau^2}e^{-z_{q,n}\tau} \;.
\ee
The integral is now divergent at $\tau=0$, but this is the usual ultraviolet divergence we encounter in the Gopakumar-Vafa integrating out calculation. 
We know that for the resolved conifold, this ultraviolet contribution can be included by simply integrating over the zero pole in the contour integral formulation. So, for that case, we can write 
\begin{eqnarray}
	& &\sum_{q,n} z_{q,n}\Big(\log z_{q,n}-1\Big) \rightarrow \nonumber \\
	& &\oint_{C_0} \frac{dw}{w} \frac{1}{1-e^{-2 \pi i w}}\frac{e^{-t_+w} + e^{-t_-w}}{w} \;.
\end{eqnarray}
The contour $C_0$ goes around the positive poles on the real axis, but also now includes the zero pole. This is exactly the same calculation that yields (the $\frac{1}{\lambda}$ part of) (\ref{polypotdeg}). So we find the same result as the D4-D2 setup, as expected from the topological string perspective. 

\section{Generalizations and Dual Emergence}
\label{sec:dualemer}

Our primary case of interest in this work has been the resolved conifold. From the perspective of emergence, the resolved conifold is very special because the ultraviolet physics can be simply included through the zero pole. In this section we consider more general settings, and frame them in a more general setting for emergence.

The general picture of emergence that we propose is similar in spirit to the Emergence proposal, but it differs qualitatively in certain ways. We denote this modified scenario of emergence as {\it Dual Emergence}. In this subsection we explain the general idea, building on the results and discussions in \cite{Hattab:2023moj,Hattab:2024thi,Hattab:2024ewk}. 

The complications for the general cases is that in general there is a phase transition between the infrared and ultraviolet regimes of the theory. The ultraviolet phase is strongly-coupled, and in general we lack a description of it. In such settings, emergence is subtle because the ultraviolet physics which is responsible for the tree-level terms in the effective action does not arise from the phase where the integrating out can be captured by considering a tower of independent particle states. In fact, it was argued in \cite{Hattab:2023moj,Hattab:2024thi,Hattab:2024ewk} that the modes which are integrated out to yield the tree-level physics are constituent degrees of freedom from which the tower of particle states are constructed as macroscopic configurations. The term Dual Emergence is capturing the point that there are two parts to the emergence: there is the ultraviolet phase from which the tree-level terms emerge, and there is the infrared phase, for which there is a description in terms of a tower of states. 

In terms of the full picture, we should consider the whole tower of states as part of a single object. Since the states being integrated out are wrapped branes, in the infrared, the ultraviolet object which captures all of them is a type of `brane field' $\psi$. Integrating out all of $\psi$ leads to the full effective action. This was studied, qualitatively and quantitatively, in \cite{Hattab:2024thi}. This was possible because in some special cases (for non-compact Calabi-Yau manifolds), there is a description of the ultraviolet phase in terms of a Fermi gas \cite{Marino:2011eh}. More precisely, the Fermi gas description is a strongly coupled description of the string theory, which we propose should describe the ultraviolet. The Fermi gas description is a quantum mechanical one with a fixed number of particles, and an associated chemical potential. In this description, the effective action, through the prepotential, is given by the Grand Canonical potential \cite{Marino:2011eh}. In \cite{Hattab:2024thi} this was explained in terms of emergence: the Grand Canonical potential is nothing but the effective action after integrating out a (second quantized) field $\psi$ of the fermions. This is precisely the $\psi$ we have discussed above. So integrating $\psi$ out generates the tree-level terms. 

In \cite{Hattab:2024thi} it was proposed how to understand this picture more generally, also for compact Calabi-Yau manifolds. The proposal, inspired by \cite{Grassi:2014zfa}, is that the classical (in the strongly-coupled description) effective action, or grand canonical potential, in terms of the chemical potential should be identified with the so-called D6 period of the Calabi-Yau (rather than the prepotential). The period has a Mellin-Barnes contour integral formulation, much like the contour integral formulation of the prepotentials we have been studying. It should be understood as arising from an integrating out calculation \cite{Hattab:2024thi}, and again the zero pole yields the tree-level terms (in terms of the chemical potential). The map from the chemical potential to the moduli fields, and the usual supergravity effective action, involves the string coupling and so is more difficult to determine. However, the leading terms, the cubic piece in the genus-zero prepotential and the linear piece in the genus-one prepotential were proposed to be extractable in \cite{Hattab:2024thi}. It would be interesting to try and similarly generalize the results of this work to compact Calabi-Yau manifolds, perhaps using some periods.
 
While more difficult to quantitatively analyse, the Dual Emergence picture does yield an answer to the question of how are the tree-level terms in the prepotential generated in cases where, because of extended supersymmetry, there is no contribution to the prepotential from integrating out half-BPS particles (say in ${\cal N}=4$ supergravity settings where there are non-renormalization results for the prepotential). This is because it is not the infrared particle tower modes of $\psi$ which yield the tree-level terms, but its ultraviolet modes that are part of a different phase of the theory and are not subject to the non-renormalization results for BPS particles in the infrared (weakly-coupled) phase. 

\section{Summary and discussion}
\label{sec:sum}

In this paper we studied integrating out M2 states ending on M5 branes and the resulting terms in the open topological string free energy or the effective supergravity action. We showed, generalizing the results of \cite{Hattab:2024ewk}, that the integrating out calculation is given by a contour integral in complexified Schwinger proper time, and that it captures non-perturbative physics.

The contour integral formulation suggests that the ultraviolet modes of the integrating out should be captured by the zero pole. We proposed that this is the case for the resolved conifold. We showed that the zero pole matches precisely the tree-level superpotential terms in the open string sector deduced through a geometric transition. This is another new check on the idea that tree-level classical terms in string theory are emergent. Interestingly, in the open string case the terms are potential terms rather than kinetic terms.

We discussed how the zero pole and ultraviolet physics behaves for more general Calabi-Yau manifolds. We proposed that the general setting for this is the scenario of Dual Emergence, described in section \ref{sec:dualemer}, where the ultraviolet contribution is in a different phase of the theory to the infrared one. This means that in general the zero pole in the contour integral of the prepotential does not capture the ultraviolet physics (but it can do so in the period).

The work presents evidence for the general idea that potentials in string theory are emergent from integrating out co-dimension one states \cite{Palti:2019pca}. While we discussed only open-string potentials in this work, we expect that the same is true for closed-string potentials. Indeed, it was already suggested in  \cite{Palti:2019pca} that type IIB flux potentials should arise from integrating out D5 branes wrapping three-cycles in a flux background. It would be very interesting to show these closed-string results in a similar way to the open-string results of this work.

In the uplift to M-theory, the D4 brane is uplifted to an M5 brane wrapping the M-theory circle, so still with a two-dimensional non-compact worldvolume. The D2 branes are uplifted to M2 branes which are not wrapping the M-theory circle but rather have momentum along it. There is also a dual picture, the particle picture discussed in \cite{Dedushenko:2014nya,Hattab:2023moj}, where one can consider the M2 branes as wrapping the M-theory circle. That wrapping becomes their worldline, and so within the non-compact dimensions they are now co-dimension two, so instantons. From this perspective, the tree-level potential is emergent from instantons. It would be interesting to develop this further, perhaps connecting to the results in \cite{Ooguri:1996me}. 

The D6 brane setting discussed in section \ref{sec:indodim1} must be implemented in a non-compact setting, such as the resolved conifold, due to sourcing tadpoles. However, one can also study similar setups with orientifold planes \cite{Sinha:2000ap}. For such setups, one should be able to implement these ideas in compact settings. 

Finally, let us comment that just as emergent kinetic terms for scalars and gauge fields are proposed to underly the Distance and Weak Gravity Conjectures \cite{Palti:2019pca}, it is natural to expect that emergent potentials underly the AdS Distance conjecture \cite{Lust:2019zwm} and the Refined de Sitter conjecture \cite{Ooguri:2018wrx}.

\vspace{0.1cm}
{\bf Acknowledgements}
\noindent
 The work of JH and EP is supported by the Israel Science Foundation (grant No. 741/20) and by the German Research Foundation through a German-Israeli Project Cooperation (DIP) grant ``Holography and the Swampland". The work of EP is supported by the Israel planning and budgeting committee grant for supporting theoretical high energy physics.

\bibliography{Higuchi}

\end{document}